\documentclass{jnst}
\usepackage[tablesfirst,notablist]{endfloat}
\usepackage{times}
\usepackage{txfonts}
\usepackage{graphicx}
\usepackage{footmisc}

\title{
EXFOR-based simultaneous evaluation for fast neutron-induced fission cross section of thorium-232
}
\author{
Vidya Devi$^{1}$
\thanks{Email: vidyathakur@yahoo.co.in},
Naohiko Otuka$^{1}$ ,
S. Ganesan$^{2}$ 
}
\inst{
$^{1}$Nuclear Data Section, 
      Division of Physical and Chemical Sciences,
      Department of Nuclear Sciences and Applications,
      International Atomic Energy Agency,
      A-1400 Wien, Austria\\
$^{2}$Former Raja Ramanna Fellow,
      Bhabha Atomic Research Centre,
      Mumbai 400085, India
}
\abst{
The $^{232}$Th neutron-induced fission cross section was evaluated from 500~keV to 200~MeV.
The experimental $^{232}$Th fission cross sections and their ratios to the $^{235,238}$U fission cross sections in the EXFOR library were reviewed and analysed by the least-squares method.
The newly published $^{232}$Th/$^{235}$U fission cross section ratios from the time-of-flight measurements at the CERN n\_TOF and CSNS Back-n facilities were compiled in EXFOR.
Additional simultaneous evaluation was performed by including the experimental $^{233,238}$U and $^{239,240,241}$Pu fission cross sections and their ratios.
The new evaluation provides the $^{232}$Th fission cross section systematically lower than the JENDL-5 cross section.
The reduction is 4\% in the plateau region between 2 and 6~MeV and more significant in the subthreshold fission region.
The present evaluation reduces the $^{232}$Th fission cross section averaged over the $^{252}$Cf spontaneous fission neutron spectrum from the JENDL-5 evaluation by 4\%, 
which is closer to the other general purpose libraries but underestimates Grundl et al's measurement by 11\%.
}
\kword{
thorium-232,
fission,
simultaneous evaluation,
JENDL,
EXFOR
}
\begin{document}
\maketitle
\section{Introduction}
Over the past three decades,
there has been renewed interest in studying the thorium fuel cycle as a potential source for new types of nuclear fuel.
In comparison to the uranium fuel cycle,
thorium-based nuclear systems are anticipated to offer a significantly higher level of intrinsic safety.
Design studies employing Th-U fuel have also shown the possibility of developing nuclear power reactors that are more resistant to nuclear proliferation than the U-Pu fuel cycle.
Additionally,
research has been reported on design systems aimed at achieving the goal of efficiently reducing weapon-grade plutonium using thorium fuel.
Another benefit of using thorium fuel is the reduced creation of higher actinide wastes.
For instance,
the Indian approach uses a closed fuel cycle using several fuels.
There are many critical reviews of thorium as a nuclear fuel with huge potential to create nuclear energy for centuries with better safety characteristics available in the literature~\cite{Revol2016Thorium,Dekoussar2005Thorium,Jyothi2023Overview}.
 
The Accelerator Driven Subcritical System (ADSS) can help to make use of thorium in power generation faster~\cite{Rubbia1995Conceptual}.
In ADSS or comparable fast reactors,
$^{232}$Th fission accounts for 3\% of the fissions and at least for 8\% in a gas cooled reactor.
Indeed,
$^{232}$Th has an important contribution of 2\% to delayed neutrons in these advanced reactor systems around the fission threshold ($\sim$0.7~MeV) as compared to 0.25\% for $^{233}$U~\cite{Abbondanno2001Measurements}.
 Therefore,
 accurate neutron-induced fission cross section of the order of 5\% is required ~\cite{Pronyaev1999Summary}.
The cross section also plays an important role in estimations of energy amplification and neutron breeding in thorium blanketed fusion-fission hybrid reactors which have significantly harder neutron spectra~\cite{Liu2018Fusion}. 

Three decades ago,
there was an increase of interest in enhancing the nuclear data of the thorium fuel cycle.
In the Indian context,
there has previously ~\cite{Ganesan2005New,Ganesan2007Nuclear,Ganesan2016Nuclear} been a brief discussion of the disparities that currently exist and the need for new and improved data. 
As multiple fuel cycles (e.g., U-Pu, Th-U) with the option of closing the fuel cycle are envisaged,
the nuclear data requirements necessary to develop new systems such as Advanced Heavy Water Reactor (AHWR)~\cite{Maheshwari2021Advanced,Vijayan2017Role} with high burnup are demanding and include the entire range of actinides and fission products for multiple fuels. 
The International Atomic Energy Agency (IAEA) also executed a Coordinated Research Project (CRP) on ``Evaluated Nuclear Data for the Th-U Fuel Cycle"~\cite{Capote2010Evaluated}.

Among the $^{232}$Th fission cross section in the fast neutron region compiled in the major general purpose libraries,
the JENDL-5 library~\cite{Iwamoto2023Japanese} adopts the JENDL-4.0~\cite{Shibata2011Japanese} cross section,
which was obtained by evaluation of the experimental cross sections (including $^{232}$Th/$^{235}$U fission cross section ratios multiplied by the JENDL-3.3 $^{235}$U fission cross section) above 300~keV by the least-squares analysis code GMA~\cite{Poenitz1997Simultaneous}.
The CENDL-3.2 library adopts the CENDL-3.1 cross section calculated by the UNF code~\cite{Han2003Calculation} but with revision between 130 and 900~keV considering the measurement by Fursov et al.~\cite{Fursov1991Measurements}.
The ENDF/B-VIII.0~\cite{Brown2018ENDF} cross section is a revision from the cross section evaluated by the EMPIRE code~\cite{Herman2007EMPIRE} for the IAEA Coordinated Research Project~\cite{Capote2010Evaluated},
and it is also adopted by the JEFF-3.3~\cite{Plompen2020Joint} and TENDL-2021~\cite{Koning2019TENDL} libraries.
The BROND-3.1~\cite{Blokhin2016New} cross section is originally from the JENDL-4.0 evaluation but adjusted to conserve the reaction cross section constrained by the optical model.
Among these evaluations, the fission cross section averaged over the $^{252}$Cf spontaneous fission neutron spectrum from JENDL-5 is closest to the experimental values from the most recent measurements by Grundl et al.~\cite{Grundl1983Fission} and Dezs\"{o} et al.~\cite{Dezso1978np},
and the JENDL-4.0 evaluation is adopted not only by the JENDL-5 library but also by the IRDFF-II library~\cite{Trkov2020IRDFF-II}.

The last measurement considered by the JENDL-4.0 evaluation is the time-of-flight measurement done by Shcherbakov et al.~\cite{Shcherbakov2002Neutron}. About 20 years after the Shcherbakov et al's measurement,
three new time-of-flight measurement results covering 1 to 200~MeV were published by the CSNS Back-n~\cite{Chen2023Measurement,Ren2023aMeasurement} and CERN n\_TOF facility~\cite{Tarrio2023Neutron} collaborations this year (2023).
Additionally, two new measurements done with monoenergetic neutron sources were published in 2021-2022~\cite{Michalopoulou2021Measurement,Gledenov2022Cross}.

The purpose of this article is to reevaluate the $^{232}$Th neutron-induced fission cross section between 500~keV and 200~MeV by including these experimental datasets newly added in the EXFOR library~\cite{Otuka2014Towards}, and to check whether the evaluation considering these new measurements support the JENDL evaluation or the cross sections in the other major general purpose libraries.

\section{Evaluation Method}
\label{sec:method}
The present evaluation was conducted using the least-squares analysis code SOK~\cite{Kawano2000Simultaneous,Kawano2000Evaluation},
wherein a prior estimate was updated with individual experimental datasets.
For $^{232}$Th, the cross section in JENDL-5 (below 20~MeV) and the cross section evaluated by Yavshits et al.~\cite{Yavshits2001Multiconfiguration} (above 20~MeV) were adopted as the prior estimates.
For other nuclides, the JENDL-5 cross sections were adopted as the prior estimate except for the $^{233}$U cross section above 20~MeV where the simultaneous evaluation result submitted to the JENDL-5 development~\cite{Otuka2022bEXFOR} was used.

We followed the JENDL-5 simultaneous evaluation procedure~\cite{Otuka2022aEXFOR} and the procedures specific for the current evaluation are briefly outlined below.
We first performed simultaneous evaluation with an experimental database compiling the $^{232}$Th,$^{235,238}$U fission cross sections including the $^{232}$Th/$^{235,238}$U cross section ratios (``three-nuclide evaluation").
Afterwards,
we repeated evaluation with the experimental database extended to $^{233}$U and $^{239,240,241}$Pu fission cross sections and their ratios (``seven-nuclide evaluation").

All experimental datasets of the $^{232}$Th cross section and its ratio to the $^{235,238}$U cross section in the EXFOR library were reviewed for selection of the datasets to be analysed.
The newly published $^{232}$Th/$^{235}$U fission cross section ratios from the time-of-flight measurements at the CERN n\_TOF and CSNS Back-n facilities were compiled in EXFOR during this evaluation,
and some EXFOR entries from older measurements were revised to include the experimental information in the least-squares analysis as much as possible.

The selection of the experimental datasets for evaluation adhered to the following criteria:
\begin{itemize}
\item including a data point between 500~keV and 200~MeV
\item reported from 1970 to the present
\item both statistical and systematic uncertainties available
\end{itemize}
The cutoff year (1970) was chosen to be consistent with the $^{233,235,238}$U and $^{239,240,241}$Pu fission cross section experimental database developed for the JENDL-5 simultaneous evaluation~\cite{Otuka2022bEXFOR} and used in the present evaluation.
The least-squares solution of the $^{232}$Th cross section near the lower (500~keV) and upper (200~MeV) energy boundaries may be affected by the experimental data points not only inside but also outside the energy range, 
and we included the experimental data points in 300--500~keV and 200--250~MeV in the present evaluation as long as the experimental dataset also has a data point between 500~keV and 200~MeV.
On the other hand, we excluded the following datasets from the present evaluation:
\begin{itemize}
\item measurement relative to a reference fission cross section but the ratio is not available in EXFOR~\cite{Blons1975Evidence,Blons1980Asymmetric,Garlea1992Integral}
\item measurement with neutrons from nuclear explosions~\cite{Muir1971Neutron}
\item numerical data not available from the authors in a tabulated form ~\cite{Lisowski1988Neutron,Konecny1972Suche}
\end{itemize}
After the selection,
no $^{232}$Th cross section dataset was left for our evaluation, and we used $^{232}$Th/$^{235,238}$U cross section ratio in EXFOR  without any modification except for the $^{232}$Th/$^{238}$U dataset published by Gledenov et al.~\cite{Gledenov2022Cross}.
Gledenov et al. obtained two datasets by exchanging the position of the $^{232}$Th and $^{238}$U (``backward" and "forward" measurements), and their results are compiled in EXFOR 32873.002.1 and 32873.003.1.
However, these measurements used the same Th and U samples and this fact makes correlation between the two datasets.
In order to take into account this correlation in the present evaluation,
we created an ad hoc EXFOR dataset (51011.002) from the two datasets and included it in our experimental database instead of EXFOR 32873.002.1 and 003.1.
Table~\ref{tab:explist1} summarizes the experimental $^{232}$Th/$^{235,238}$U datasets used in the current evaluation.
\begin{table}
\caption{
$^{232}$Th/$^{235,238}$U fission cross section ratios included in the experimental database for the present evaluation.
``Ver.", ``Lab." and ``Pts." give the date (N2) of the SUBENT record in EXFOR, EXFOR/CINDA abbreviation~\cite{Otuka2023EXFOR} of the institute where the experiment was performed, and number of data points, respectively.
}
\label{tab:explist1}
\begin{tabular}{lllllrccc} 
\hline
EXFOR $\#$       &Ver.    &First author    &Year&Lab.   &Pts.&\multicolumn{2}{c}{Energy range (eV)}&Ref.          \\
\hline                                                                                                           
$^{232}$Th/$^{235}$U\\                                                                                     
32889.002.1      &20230609&Z.Ren           &2023&3CPRIHP&  86&1.0E+06&2.1E+08&\cite{Ren2023aMeasurement}          \\
23654.002.1      &20230502&D.Tarr\'{i}o    &2023&2ZZZCER& 181&1.0E+06&2.5E+08&\cite{Tarrio2023Neutron}            \\
32887.002.1      &20230401&Y.Chen          &2023&3CPRIHP& 109&1.0E+06&2.4E+08&\cite{Chen2023Measurement}          \\
23756.002.1$^{a}$&20221007&V.Michalopoulou &2021&2GRCATH&   2&2.0E+06&2.5E+06&\cite{Michalopoulou2021Measurement} \\
41455.006        &20220914&O.Shcherbakov   &2002&4RUSLIN& 166&5.8E+05&2.0E+08&\cite{Shcherbakov2002Neutron}       \\
41111.002        &20221011&B.Fursov        &1991&4RUSFEI&  66&3.5E+05&7.4E+06&\cite{Fursov1991Measurements}       \\
13134.003.1      &20170724&J.W.Meadows     &1988&1USAANL&   1&1.5E+07&1.5E+07&\cite{Meadows1988Fission}           \\
22282.002.1      &20130924&F.Manabe        &1988&2JPNTOH&   4&1.3E+07&1.5E+07&\cite{Manabe1988Measurements}       \\
22014.002        &20210315&K.Kanda         &1986&2JPNTOH&  17&1.5E+06&6.8E+06&\cite{Kanda1986Measurement}         \\
40885.002        &20221017&A.A.Goverdovskii&1986&4RUSFEI&   1&1.6E+07&1.6E+07&\cite{Goverdovskii1986aMeasurement} \\
40888.002        &20221017&A.A.Goverdovskii&1986&4RUSFEI&  33&4.9E+06&1.0E+07&\cite{Goverdovskii1986bMeasurement} \\
30813.011        &20190722&I.Garlea        &1984&3RUMBUC&   1&1.5E+07&1.5E+07&\cite{Garlea1984Measuring}          \\
10843.003.1$^{b}$&20221018&J.W.Meadows     &1983&1USAANL&  69&1.3E+06&9.9E+06&\cite{Meadows1983Fission}           \\
10658.002        &20221020&J.W.Behrens     &1982&1USALRL& 144&7.1E+05&3.3E+07&\cite{Behrens1982Measurement}       \\
20844.002        &20221024&C.Nordborg      &1978&2SWDUPP&  23&4.6E+06&8.8E+06&\cite{Nordborg1978Fission}          \\
$^{232}$Th/$^{238}$U\\                              
23756.002.2$^{c}$&20221007&V.Michalopoulou &2021&2GRCATH&   2&2.0E+06&2.5E+06&\cite{Michalopoulou2021Measurement} \\
23756.003.1      &20221007&V.Michalopoulou &2021&2GRCATH&  10&3.0E+06&1.8E+07&\cite{Michalopoulou2021Measurement} \\
51011.002$^{d}$  &20221111&Yu.M.Gledenov   &2022&3CPRBJG&  24&4.2E+06&1.2E+07&\cite{Gledenov2022Cross}            \\
\hline
\end{tabular}
\begin{flushleft}
{\small
$^a$EXFOR \# is 51012.002 in our experimental database.\\
$^b$treated as shape ratio.\\
$^c$EXFOR \# is 51012.003 in our experimental database.\\
$^d$ad hoc EXFOR entry prepared from 32873.002.1 and 003.1.\\
}
\end{flushleft}
\end{table}

For the target nuclides other than $^{232}$Th,
the energy range included in fitting is the same as the JENDL-5 simultaneous evaluation,
namely from 10~keV (fissile) or 100~keV (non-fissile) to 200~MeV.
The experimental database developed for the JENDL-5 simultaneous evaluation \cite{Otuka2022bEXFOR} was adopted for these nuclides with exclusion of one $^{235}$U dataset (see the reason outlined in \cite{Otuka2023Simultaneous}) and inclusion of a few additional datasets as summarized in Table~\ref{tab:explist2}.

Belloni et al.~\cite{Belloni2022Neutron} determined the $^{238}$U cross sections (EXFOR 23653.004.1 and 23653.004.2),
which are by-products of their measurements of the $^{240,242}$Pu cross sections relative to n-p scattering and to $^{238}$U fission.
We merged these two datasets into an ad hoc EXFOR dataset 51014.002 to include in our experimental database the correlation between the two datasets originated from the use of the same neutron flux.
Note that they do not report the $^{240}$Pu/$^{238}$U ratio from their measurement, 
and we did not use their $^{240}$Pu cross section derived from the measured ratio multiplied by the $^{238}$U cross section in IAEA Neutron Data Standard~\cite{Carlson2018Evaluation} following our data selection criteria.
\begin{table}
\caption{
Update of the experimental database for $^{233,235,238}$U and $^{239,240,241}$Pu fission cross sections and their ratios
from the database used for JENDL-5 simultaneous evaluation~\cite{Otuka2022bEXFOR}.
EXFOR 31833.002 was deleted whereas the other ones were added.
See the caption of Table~\ref{tab:explist1} for ``Ver.", ``Lab." and ``Pts.".
}
\label{tab:explist2}
\begin{tabular}{llllllrccc} 
\hline
Reaction            &EXFOR $\#$ &Ver.    &First author&Year&Lab.   &Pts.&\multicolumn{2}{c}{Energy range (eV)}&Ref.   \\
\hline
$^{233}$U/$^{235}$U &23654.003.1&20230502&D.Tarr\'{i}o&2023&2ZZZCER& 110&1.0E+04&2.3E+08&\cite{Tarrio2023Neutron}     \\
$^{235}$U           &31833.002  &20201103&R.Arlt      &1980&2GERZFK&   1&8.2E+06&8.2E+06&\cite{Arlt1981Absolute}      \\
$^{235}$U           &23294.004  &20211216&I.Duran     &2019&2ZZZCER&   8&7.5E+03&2.7E+04&\cite{Duran2019High}         \\
$^{238}$U           &51014.002  &20221222&F.Belloni   &2022&2GERPTB&   4&2.5E+06&1.5E+07&\cite{Belloni2022Neutron}    \\
$^{238}$U/$^{235}$U &32886.003.1&20230206&Z.Ren       &2023&3CPRIHP& 135&5.1E+05&1.8E+08&\cite{Ren2023bMeasurement}   \\
$^{239}$Pu/$^{235}$U&14721.002  &20211220&L.Snyder    &2021&1USALAS& 119&1.1E+05&9.7E+07&\cite{Snyder2021Measurement} \\
$^{240}$Pu          &23653.002.1&20221126&F.Belloni   &2022&2GERPTB&   2&2.5E+06&1.5E+07&\cite{Belloni2022Neutron}    \\
\hline
\end{tabular}
\end{table}

The SOX code~\cite{Otuka2022bEXFOR} was utilized to convert all numerical values from the EXFOR library into the SOK input files.
In order to construct the energy-energy correlation coefficients, it was assumed during processing by SOX that the uncertainty arising from counting statistics is uncorrelated while the other uncertainties are fully correlated within the same dataset.
We assumed any two datasets in our experimental database are independent of each other.

\section{Result and Discussion}
We performed fitting to a total of 8695 (3268) experimental data points by adjusting 597 (261) evaluated cross sections as fitting parameters,
and obtained the least-squares solutions with the reduced chi-square of 4.05 (4.49) for the seven (three)-nuclide evaluation.\footnote{The evaluated cross sections in an ASCII file are available as a Supplemental Material of this article.}
The seven-nuclide evaluation shows better consistency between the fitting result and experimental data points.
%

\subsection{Point-wise cross sections}
\begin{figure}
\centering
\includegraphics[bb=0 0 842 595,trim=55 0 0 0,clip,width=1.0\textwidth]{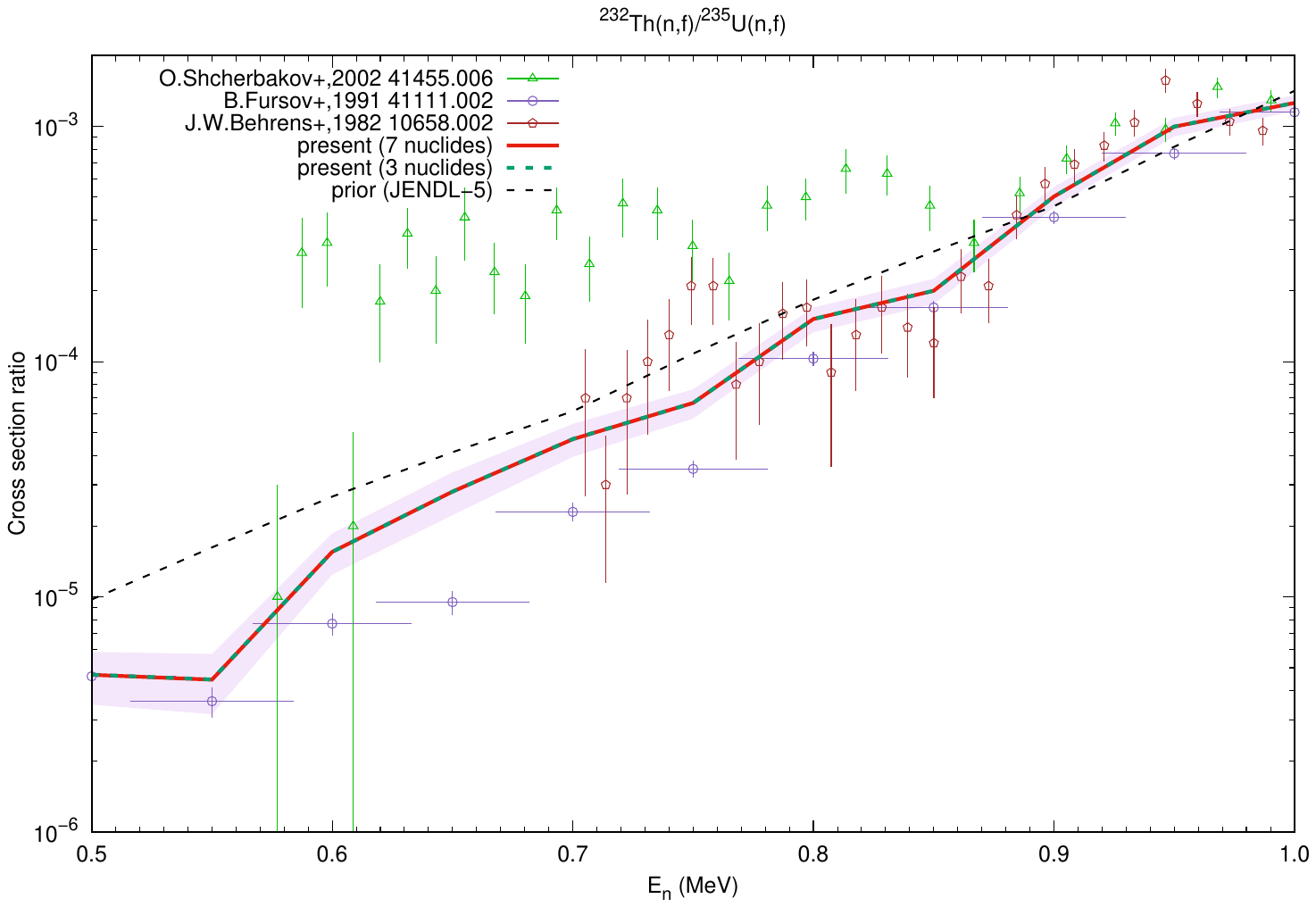}
\caption{
$^{232}$Th/$^{235}$U fission cross section ratios below 1~MeV from the present evaluations along with the experimental ratios~\cite{Shcherbakov2002Neutron,Fursov1991Measurements,Behrens1982Measurement} used in the present evaluation.
\label{fig:th232u235-low}
}
\end{figure}

\begin{figure}
\centering
\includegraphics[bb=0 0 842 595,trim=55 0 0 0,clip,width=1.0\textwidth]{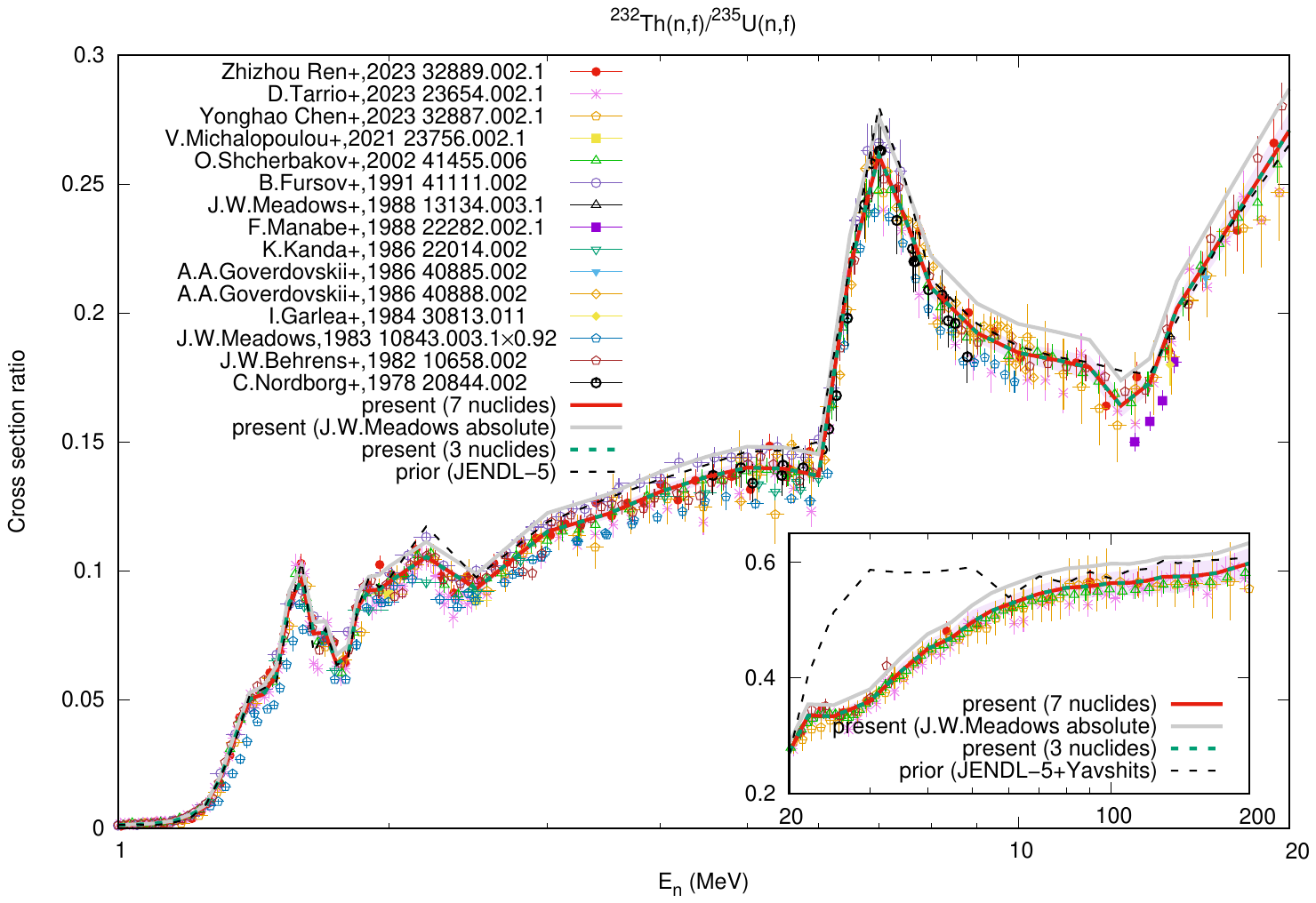}
\caption{
$^{232}$Th/$^{235}$U fission cross section ratios above 1~MeV from the present evaluations along with the experimental ratios~\cite{Ren2023aMeasurement,Tarrio2023Neutron,Chen2023Measurement,Michalopoulou2021Measurement,Shcherbakov2002Neutron,Fursov1991Measurements,Meadows1988Fission,Manabe1988Measurements,Kanda1986Measurement,Goverdovskii1986aMeasurement,Goverdovskii1986bMeasurement,Garlea1984Measuring,Meadows1983Fission,Behrens1982Measurement,Nordborg1978Fission} used in the present evaluation.
``J.W.Meadows absolute" stands for the seven-nuclide evaluation including Meadows' ratio as the absolute ratio.
}
\label{fig:th232u235-hig}
\end{figure}

\begin{figure}
\centering
\includegraphics[bb=0 0 595 842,trim=0 50 0 0,clip,width=1.0\textwidth]{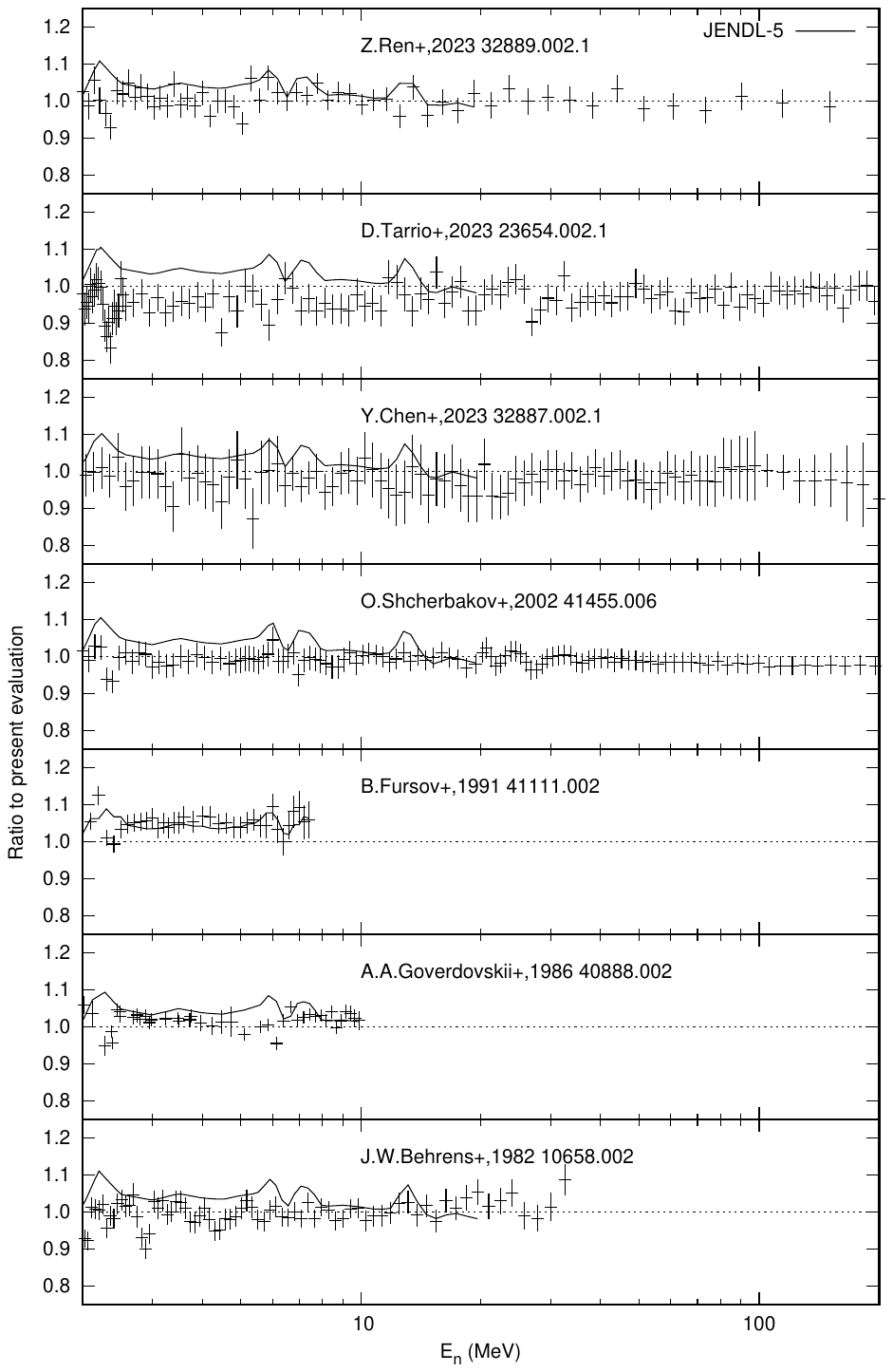}
\caption{
$^{232}$Th/$^{235}$U fission cross section ratios from time-of-flight measurements~\cite{Ren2023aMeasurement,Tarrio2023Neutron,Chen2023Measurement,Shcherbakov2002Neutron,Fursov1991Measurements,Goverdovskii1986bMeasurement,Behrens1982Measurement} relative to the ratio from present evaluation above 2~MeV.
The smooth curve shows the JENDL-5 $^{232}$Th/$^{235}$U cross section ratio relative to the newly evaluated $^{232}$Th/$^{235}$U cross section ratio.
}
\label{fig:exp-sok-rat}
\end{figure}

\begin{figure}
\centering
\includegraphics[bb=0 0 842 595,trim=55 0 0 0,clip,width=1.0\textwidth]{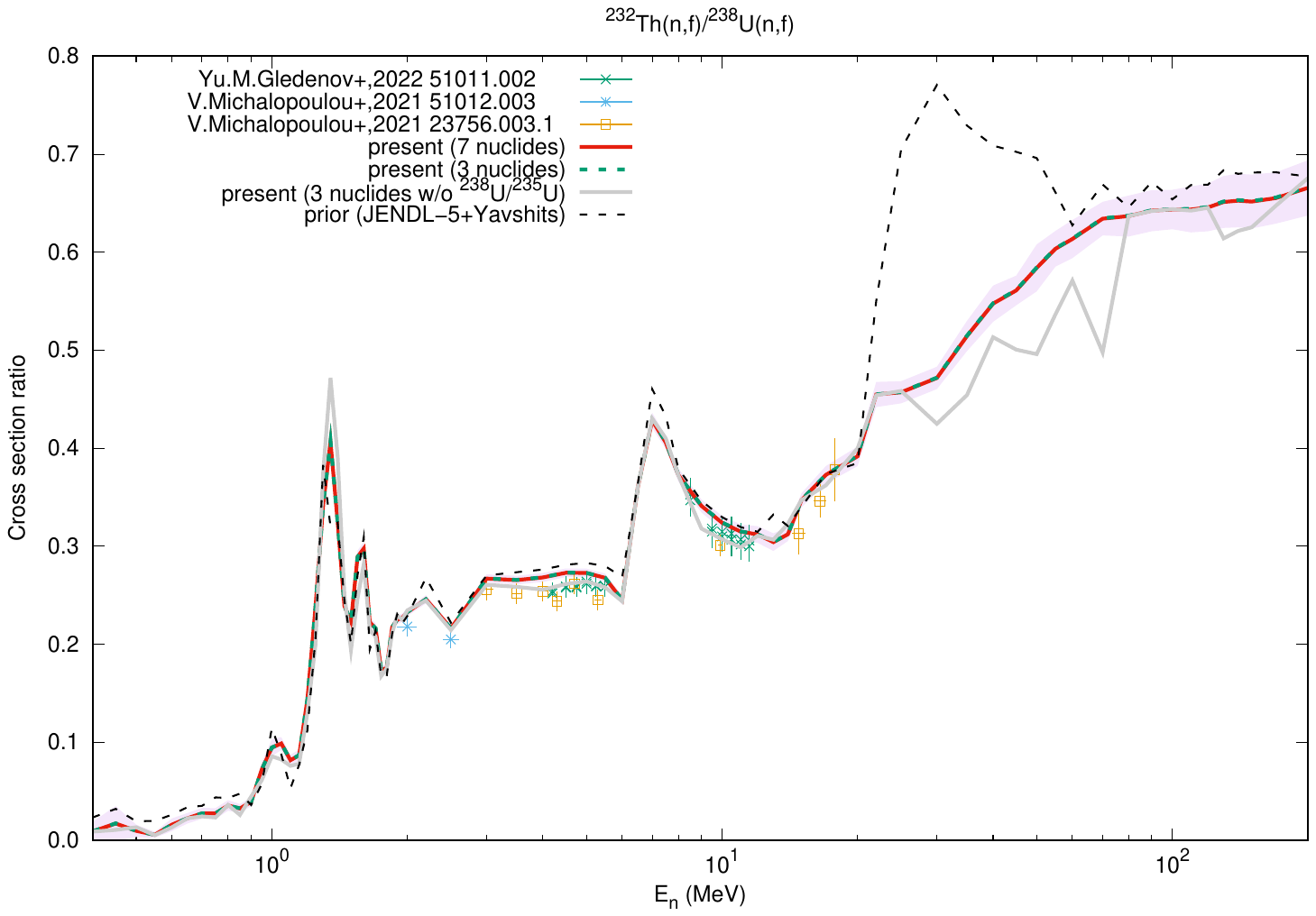}
\caption{
$^{232}$Th/$^{238}$U fission cross section ratios from the present evaluations along with the experimental ratios~\cite{Michalopoulou2021Measurement,Gledenov2022Cross} used in the present evaluation.
``3 nuclides w/o $^{238}$U/$^{235}$U" stands for the three-nuclide evaluation excluding the experimental $^{238}$U/$^{235}$U ratios.
}
\label{fig:th232u238}
\end{figure}

\begin{figure}
\centering
\includegraphics[bb=0 0 842 595,trim=55 0 0 0,clip,width=1.0\textwidth]{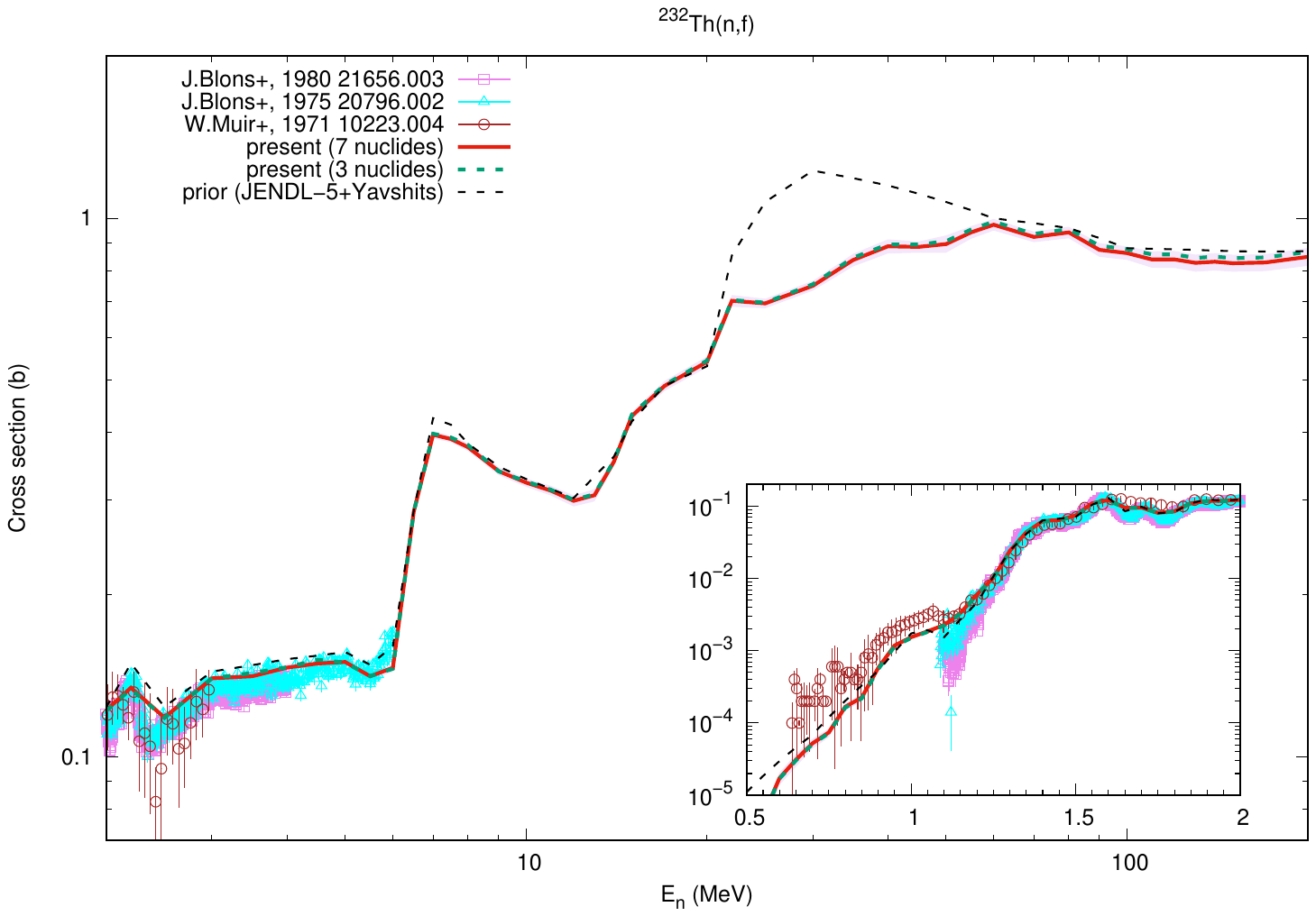}
\caption{
$^{232}$Th fission cross sections from the present evaluations along with the experimental cross sections~\cite{Blons1975Evidence,Blons1980Asymmetric,Muir1971Neutron} that were excluded from the present evaluation.
}
\label{fig:th232}
\end{figure}

The newly evaluated $^{232}$Th cross section and its ratio to the $^{235}$U and $^{238}$U cross sections are presented in Figs.~\ref{fig:th232u235-low} to \ref{fig:th232}.
The experimental data points used in the current evaluation are represented by symbols in these figures except for Fig.~\ref{fig:th232},
where none of the plotted experimental data points are used in the present evaluation and they are shown just for comparison with the result of the present evaluation. The band accompanying the newly evaluated cross section or its ratio in the figures represents the external uncertainty, which is the internal uncertainty in the least-squares solution multiplied by the square root of the reduced chi-square from the seven-nuclide evaluation.\footnote{A reduced chi-square value greater than 1 indicates that the uncertainty resulting from the least-squares fitting (internal uncertainty) is less than that anticipated from the actual discrepancy of the experimental datasets (external uncertainty). See also Sect. 4.4 of Ref. \cite{Mughabghab2018Atlas}.}

Figure~\ref{fig:th232u235-low} displays the $^{232}$Th/$^{235}$U cross section ratio below 1~MeV. The ratio is the same for both the current seven- and three-nuclide evaluations. Only three experimental datasets are available in this energy range. Among them, Fursov et al.~\cite{Fursov1991Measurements} and Behrens et al.~\cite{Behrens1982Measurement} are close to the present evaluation while Shcherbakov et al.~\cite{Shcherbakov2002Neutron} do not agree with these two datasets within their respective error bars below 0.85~MeV. Considering low counting statics of $^{232}$Th fission,
Fursov et al.~\cite{Fursov1991Measurements} changed the fission fragment detection device at 0.85~MeV from an ionization chamber to a track (mica) detector and it could give a more reasonable estimation of the actual ratio than Shcherbakov et al.
In presence of Shcherbakov et al's high ratio,
the ratio from the present evaluation becomes slightly higher than Fursov et al's ratio below 0.7~MeV.
Between 0.85 and 1~MeV,
all three experimental datasets come closer and show better consistency with the JENDL-5 and present evaluations.

Figure~\ref{fig:th232u235-hig} displays the $^{232}$Th/$^{235}$U cross section ratio above 1~MeV.
The ratio initially obtained from the seven-nuclide evaluation was systematically higher than JENDL-5 and obviously did not represent the experimental information.
We found out that this problem is solved by treating Meadows' ratio~\cite{Meadows1983Fission} as the shape ratio (i.e., treat its overall normalization factor as an additional fitting parameter).
This treatment gives more reasonable fitting with 0.922$\pm$0.010 (external uncertainty) as the normalization factor to be multiplied to the original Meadows' ratio.
We also attempted fitting excluding Meadows' ratio and confirmed that the result is very similar to the fitting with Meadows' ratio as a shape ratio.
This figure shows that the present ratio is systematically lower than the JENDL-5 ratio from 2 to 6~MeV and 7 to 15~MeV.
In the plateau between 2 and 6~MeV, JENDL-5 follows the systematically high ratio measured by Fursov et al.~\cite{Fursov1991Measurements} while the present fitting does not support it.
The inset of Fig.~\ref{fig:th232u235-hig} shows that the $^{232}$Th/$^{235}$U ratio in the high energy region derived from the $^{235}$U cross section in JENDL-5 and $^{232}$Th cross section evaluated by Yavshits et al. is too high while the present evaluation provides a ratio consistent with the experimental datasets.
In the high energy region,
the two ratios from CSNS Back-n~\cite{Chen2023Measurement} and CERN n\_TOF~\cite{Tarrio2023Neutron} published this year (2023) are consistent with the ratio published by Shcherbakov et al's ratio.
Some data points from the CERN n\_TOF measurement are lower than but still consistent with others within the error bars.
The present seven- and three-nuclide evaluations provide almost the same ratio above 1~MeV.

Figure~\ref{fig:exp-sok-rat} shows the experimental $^{232}$Th/$^{235}$U cross section ratio relative to the newly evaluated $^{232}$Th/$^{235}$U cross section ratio for recent time-of-flight measurements.
Additionally, the JENDL-5 $^{232}$Th/$^{235}$U cross section ratio relative to the newly evaluated $^{232}$Th/$^{235}$U cross section ratio is shown by smooth curve.
This figure shows the JENDL-5 evaluation follows the ratio measured by Fursov et al. very well.
The JENDL-5 evaluation also considers the ratio measured by Shcherbakov et al., Goverdovskii et al. and Behrens et al.,
and it is not clear why the JENDL-5 evaluation is influenced by the ratio measured by Fursov et al.
It is obvious from this figure that the newly evaluated $^{232}$Th/$^{235}$U ratio is more consistent with the majority of the time-of-flight measurements published before and after JENDL-5 evaluation.
Among the three datasets published in 2023, only Ren et al.~\cite{Ren2023aMeasurement} determined the overall normalization experimentally.
Tarr\'{i}o et al.~\cite{Tarrio2023Neutron} normalized their shape ratio to an absolute ratio at 2.5 - 5 MeV from a previous n\_TOF measurement, and Chen et al.~\cite{Chen2023Measurement} normalized their shape ratio to 0.364$\pm$0.007~b at 14.13~MeV.
%

Figure~\ref{fig:th232u238} shows the $^{232}$Th/$^{238}$U ratio.
There are three new experimental datasets measured with monoenergetic neutron sources published in 2021-2022~\cite{Gledenov2022Cross,Michalopoulou2021Measurement}.
The ratio from the present evaluation is systematically lower than the JENDL-5 ratio between 2 and 15~MeV and more consistent with but still systematically higher than the newly measured ratios.
In order to understand this deviation,
we performed additional fitting for the three-nuclide evaluation excluding the $^{238}$U/$^{235}$U experimental ratios,
and we found this exclusion improves the agreement with the evaluated and measured $^{232}$Th/$^{238}$U ratios a lot as shown in Fig.~\ref{fig:th232u238}.
This fact reminds us that the scale of the $^{238}$U cross section from the JENDL-5 simultaneous evaluation including the $^{238}$U/$^{235}$U experimental ratios is about 3.5\% lower than the scale expected from the $^{252}$Cf fission neutron spectrum averaged cross section measured by Grundl et al.~\cite{Grundl1983Fission} and recommended by Mannhart~\cite{Mannhart2006Response}.

\begin{figure}
\centering
\includegraphics[bb=0 0 842 595,trim=55 0 0 0,clip,width=1.0\textwidth]{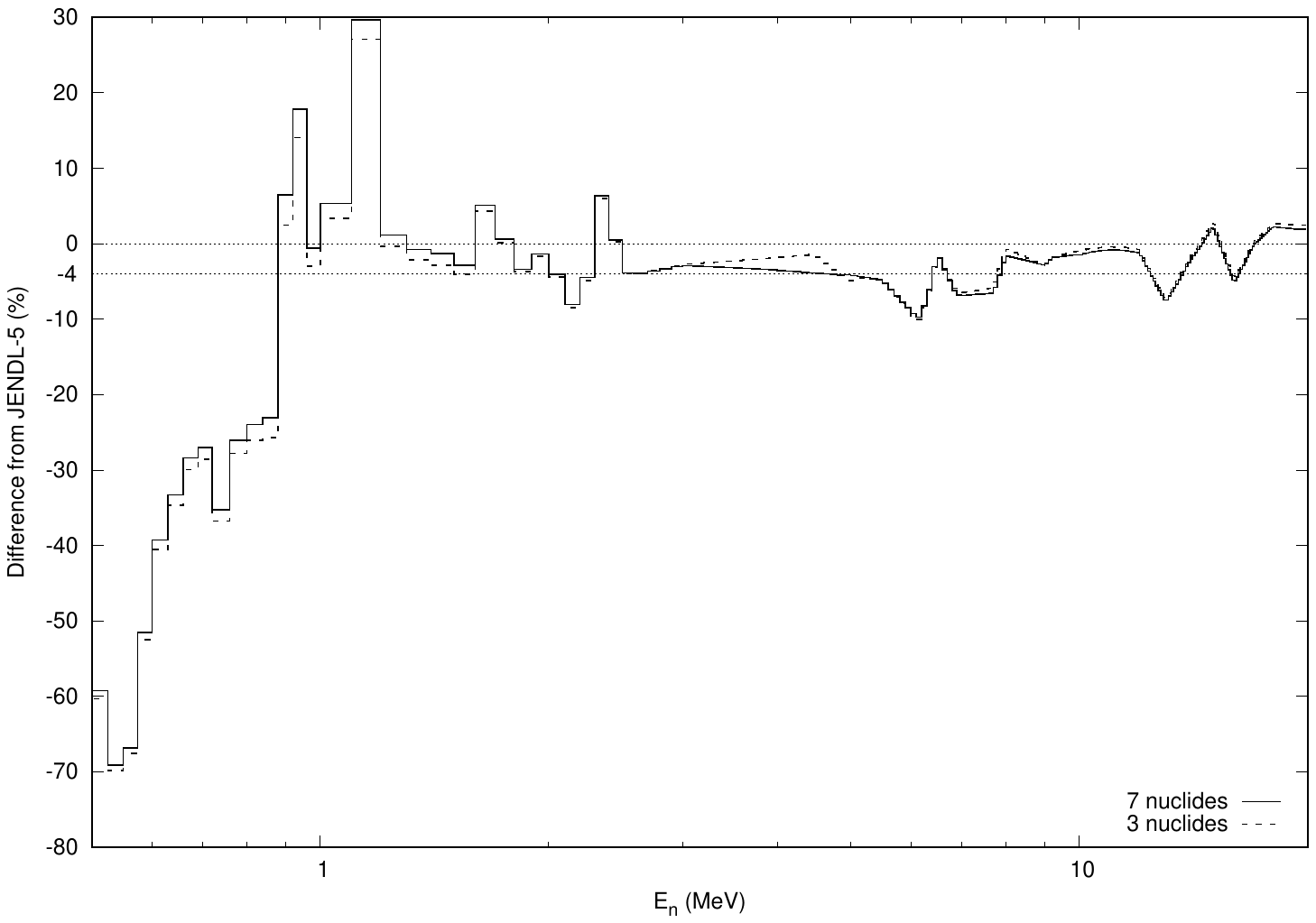}
\caption{
Difference in $^{232}$Th fission cross sections between evaluations.
}
\label{fig:histo}
\end{figure}

Figure~\ref{fig:th232} shows that the $^{232}$Th cross section from the current evaluation along with the three datasets~\cite{Blons1980Asymmetric,Blons1975Evidence,Muir1971Neutron} that were published later than 1970 but do not fulfill the conditions mentioned in Sect.~\ref{sec:method} and excluded from the present evaluation.
Muir et al.~\cite{Muir1971Neutron} determined the cross section by using the measured $^{232}$Th/$^{239}$Pu ratio with the $^{239}$Pu cross section evaluated by Davey~\cite{Davey1968Selected}.
Muir et al.'s cross section below 1~MeV is systematically higher than the cross section from the current evaluation.
The two datasets of the cross section measured by Blons et al.~\cite{Blons1975Evidence,Blons1980Asymmetric} show good agreement with the cross section from the present evaluation except for the energy region near the upper boundary of their measurement.
Figure~\ref{fig:histo} illustrates the change in the $^{232}$Th cross sections in the new evaluations from the JENDL-5 evaluation in the SAND-II 725 energy group structure.
The figure shows that the cross sections from the three- and seven-nuclide evaluations are very similar and they are systematically lower than the JENDL-5 cross section except for a few groups.
The reduction from the JENDL-5 cross section is about 4\% in the plateau region between 2 and 6~MeV.
This partly resolves the discrepancy between the JENDL-5 and n\_TOF cross section (7\%) discussed in Ref.~\cite{Tarrio2023Neutron}.
Below 1~MeV, the deviations from the JENDL-5 evaluation become more significant as it is also seen in Fig.~\ref{fig:th232u235-low}.

\begin{figure}
\centering
\includegraphics[bb=0 0 842 595,trim=55 0 0 0,clip,width=1.0\textwidth]{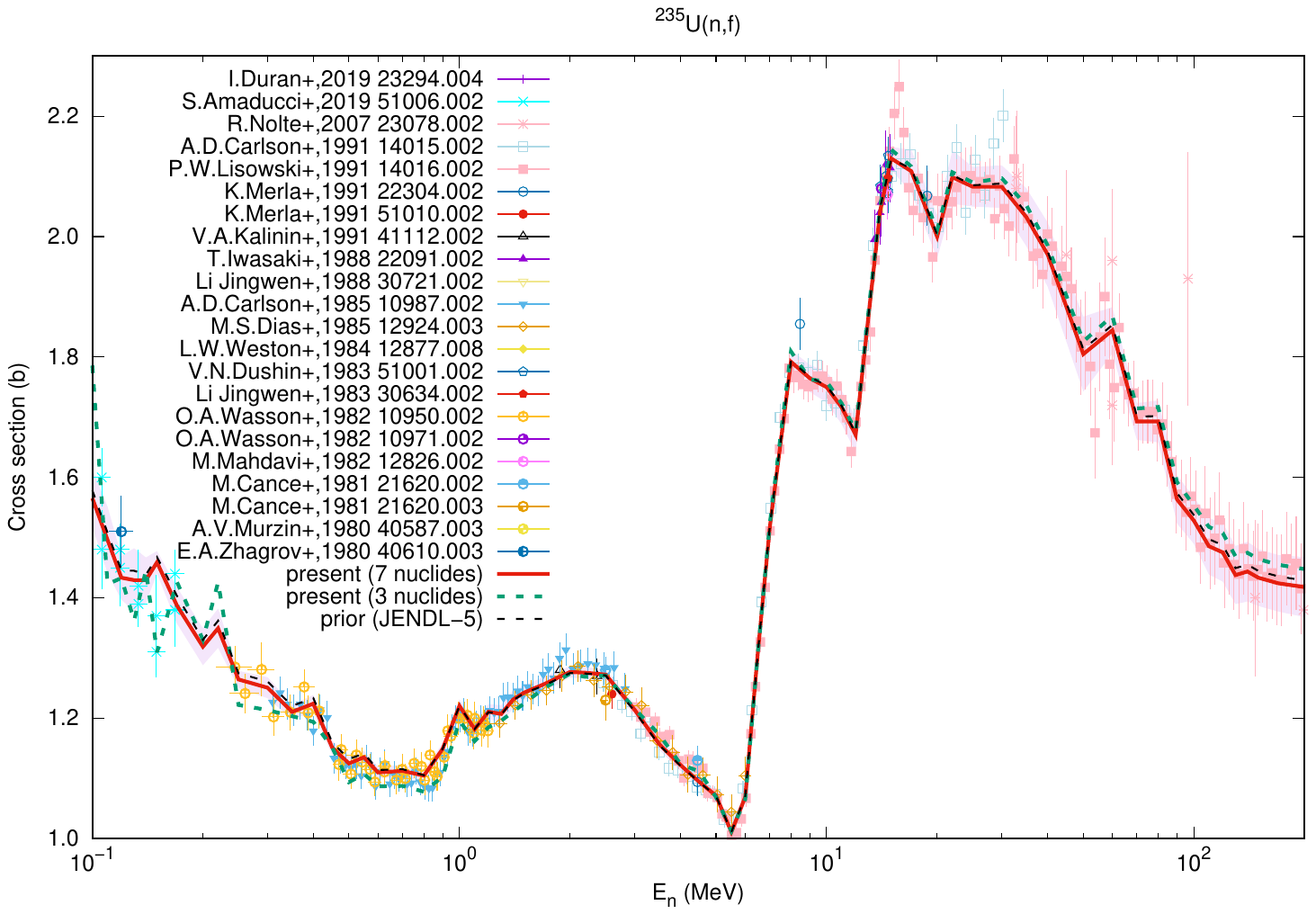}
\caption{
$^{235}$U fission cross sections from the present evaluations along with the experimental cross sections~\cite{Duran2019High,Amaducci2019Measurement,Nolte2007Cross,Carlson1992Measurements,Lisowski1991Fission,Merla1991Absolute,Kalinin1991Correction,Iwasaki1988Measurement,Li1988Absolute,Carlson1985Absolute,Dias1985Application,Dushin1983Statistical,Li1983Absolute,Wasson1982Absolute,Wasson1982Measurement,Weston1984Subthreshold,Mahdavi1983Measurements,Cance1981Measures,Murzin1980Measurement,Zhagrov1980Fission} used in the present evaluation.
} 
\label{fig:u235}
\end{figure}

The present evaluation is simultaneous evaluation and the least-squares fitting provides not only the $^{232}$Th cross section but also the $^{233,235,238}$U and $^{239,240,241}$Pu ($^{235,238}$U) cross sections in the seven-nuclide (three-nuclide) evaluation.
Among these cross sections,
the $^{235}$U cross section adopted in the JENDL-5 library has been validated by various integral benchmark tests (e.g., ~\cite{Tada2023JENDL-5}) and the cross section from the present evaluation should be very close to the JENDL-5 cross section.
Figure~\ref{fig:u235} compares the $^{235}$U cross section from the present evaluation with the prior cross section taken from the JENDL-5 library.
The figure reveals that the seven-nuclide evaluation does not introduce significant changes from the JENDL-5 $^{235}$U cross section below 100~MeV.
Conversely, the three-nuclide evaluation systematically reduces the prior estimate (i.e., JENDL-5 cross section) between 0.2 and 2~MeV.
The unphysical structure of the cross section between 0.1 and 0.2~MeV from the three-nuclide evaluation is resolved in the seven-nuclide evaluation.
Based on these findings,
we conclude that the seven-nuclide evaluation provides a more reasonable assessment of the $^{232}$Th cross section though the two evaluations do not show much difference as seen in Fig.~\ref{fig:histo}.

\subsection{Validation by $^{252}$Cf spontaneous fission neutron spectrum averaged cross section}
The $^{252}$Cf spontaneous fission neutron field over the energy range from 0.1 to 20~MeV is recognized as the standard neutron field by the IRDFF-II library.
The JENDL-5 simultaneous evaluation~\cite{Otuka2022aEXFOR} utilized the spectrum averaged cross section (SACS) measured in this neutron field for validation of the evaluated $^{235,238}$U and $^{239}$Pu cross sections by comparison with those recommended by Mannhart~\cite{Mannhart2006Response}.
He does not provide a recommended value for the $^{232}$Th cross section but quotes in the same document~\cite{Mannhart2006Response} the SACS measured by Grundl et al.~\cite{Grundl1983Fission} (89.4$\pm$2.7~mb, the latest experimental $^{252}$Cf SACS of the $^{232}$Th cross section in EXFOR) for comparison with the SACS derived from the $^{232}$Th cross sections in two dosimetry libraries (IRDF-90.2~\cite{Kocherov1996International} and JENDL/D-99~\cite{Kobayashi2002JENDL}).
To validate the newly evaluated $^{232}$Th cross sections cross sections,
we converted the evaluated point-wise cross sections into group-wise cross sections within the SAND-II 725 energy group structure,
and averaged over the $^{252}$Cf spontaneous fission neutron spectrum in the same group structure compiled in the IRDFF-II library following the procedure described in Ref.~\cite{Otuka2022aEXFOR}.
The SACS of the present results was obtained by joining the newly evaluated cross section above 500 keV to the cross section in the JENDL-5 library below 500 keV.

\begin{figure}
\centering
\includegraphics[bb=0 0 842 595,trim=55 0 0 0,clip,width=1.0\textwidth]{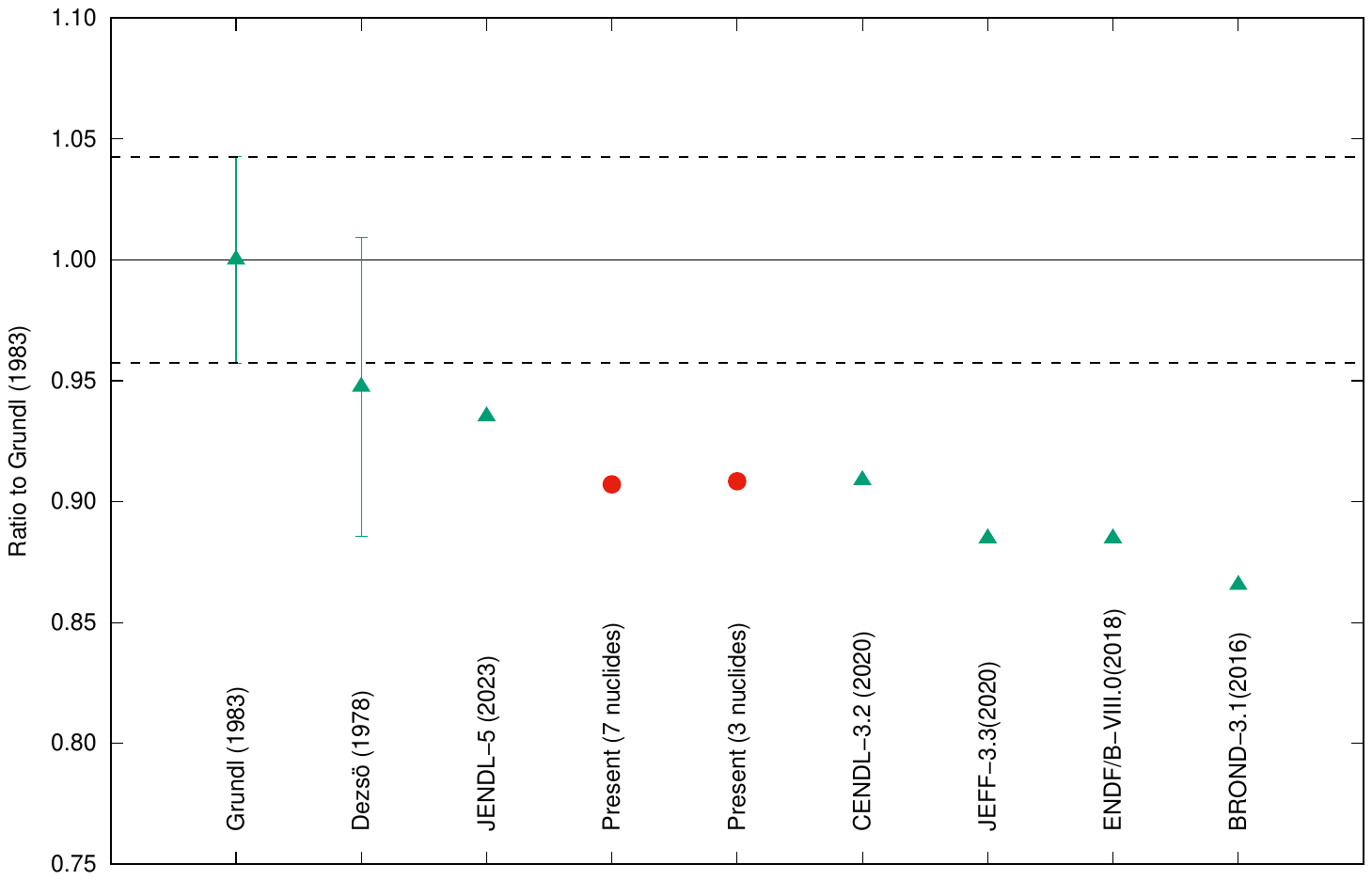}
\caption{
$^{252}$Cf spontaneous fission neutron spectrum averaged cross sections relative to those measured by Grundl et al.~\cite{Grundl1983Fission} for the present evaluation, evaluated data libraries~\cite{Iwamoto2023Japanese,Ge2020CENDL,Plompen2020Joint,Blokhin2016New,Brown2018ENDF} as well as the one measured by Dezs\"{o}~\cite{Dezso1978np}.
The JEFF-3.3 adopts ENDF/B-VIII evaluation.
}
\label{fig:cf252sacs}
\end{figure}

Figure~\ref{fig:cf252sacs} shows the ratios to the SACS measured by Grundl et al. for the present evaluation, JENDL-5, CENDL-3.2, JEFF-3.3, ENDF/B-VIII.0 and BROND-3.1 evaluations.
Note that the JEFF-3.3 library adopts the ENDF/B-VIII.0 $^{232}$Th cross section.
The figure shows that the SACS from all evaluations are inconsistent with the SACS measured by Grundl et al. despite the fact that all SACS from these evaluations other than BROND-3.1 evaluation fall within the error bar of the second most recent experimental SACS measured by  Dezs\"{o} et al.~\cite{Dezso1978np}. 
The present update from JENDL-5 increases the discrepancy from the measurement by Grundl et al. from 7\% to 11\%.

Table~\ref{tab:cf252sacs} summarizes the SACS from the present evaluations, JENDL-5 simultaneous evaluation (sok20210404)~\cite{Otuka2022aEXFOR} and its update (sok20220324)~\cite{Otuka2023Simultaneous} in parallel with the measured, evaluated and recommended values.
Note that the SACS derived from the JENDL-5 library is slightly different from the SACS from the simultaneous evaluation submitted to the JENDL project (sok20210404) due to adjustments~\cite{Iwamoto2023Japanese,Otuka2022bEXFOR}.
The SACS from the present evaluation is 4\% lower than the $^{232}$Th SACS in JENDL-5,
which is consistent with what we observe in Fig.~\ref{fig:histo}.
Table 3 shows that the $^{233,235,238}$U and $^{239,240,241}$Pu SACS from the present evaluation is almost unchanged from those derived in the latest SOK simultaneous fitting~\cite{Otuka2023Simultaneous} except for the $^{235}$U SACS from the three-nuclide evaluation.
\begin{table}
\caption{
Californium-252 spontaneous fission neutron spectrum averaged cross sections (mb).
}
\label{tab:cf252sacs}
\begin{tabular}{llllllll} 
\hline
                                           &$^{232}$Th  &$^{233}$U  &$^{235}$U  &$^{238}$U     &$^{239}$Pu  &$^{240}$Pu &$^{241}$Pu   \\
\hline                                                                                        
Present (7 nuclides)                       &81.1        &1900       &1220       &314           &1812        &1337       &1606         \\
Present (3 nuclides)                       &81.2        &           &1209       &314           &            &           &             \\
sok20220324~\cite{Otuka2023Simultaneous}   &            &1899       &1221       &315           &1814        &1339       &1608         \\
sok20210404~\cite{Otuka2022aEXFOR}         &            &1900       &1223       &316           &1808        &1340       &1606         \\
JENDL-5~\cite{Iwamoto2023Japanese}         &83.6        &1901       &1221       &321           &1808        &1340       &1606         \\
Dezs\"{o} and Csikai\cite{Dezso1978np}     &84.7$\pm$4.9&           &           &310.9$\pm$14.0&            &           &             \\
Grundl and Gilliam~\cite{Grundl1983Fission}&89.4$\pm$2.7&1893$\pm$48&1216$\pm$19&326$\pm$6.5   &1824$\pm$35 &1337$\pm$32&1616$\pm$80  \\
Mannhart~\cite{Mannhart2006Response}       &            &           &1210$\pm$15&325.7$\pm$5.3 &1812 $\pm$25&           &             \\
\hline
\end{tabular}
\end{table}
\section{Summary}
We presented a new evaluation of the $^{232}$Th neutron-induced fission cross section by the least-squares method in the energy range from 500~keV to 200~MeV.
The simultaneous evaluation considers not only the $^{232}$Th but also the $^{235,238}$U fission cross sections to include the experimental $^{232}$Th/$^{235}$U and $^{232}$Th/$^{238}$U fission cross section ratios in the evaluation.
We also expanded the scope of the evaluation from the three nuclides to seven nuclides including the $^{233}$U and $^{239,240,241}$Pu fission cross sections and their corresponding ratios in fitting.

The three- and seven-nuclide evaluations give similar $^{232}$Th neutron fission cross sections, and they are systematically lower than the JENDL-5 cross sections.
The reduction is about 4\% in the plateau region between 2 and 6~MeV and more significant in the subthreshold region.
Comparisons with the measured and evaluated values were done for the $^{252}$Cf fission neutron spectra averaged cross section (SACS) of the $^{232}$Th fission cross section.
The SACS from the present evaluation is lower than the SACS measured by Grundl et al. by 11\% while closer than the JENDL-5 SACS to the SACS of the other general purpose libraries.
%
\section*{Acknowledgement}
We are grateful to Yonghao Chen (Institute of High Energy Physics, China), Diego Tarr\'{i}o (Uppsala University, Sweden) and Zhizhou Ren (University of Science and Technology of China, China) for sharing with us the newly published $^{232}$Th/$^{235}$U ratio data and Veatriki Michalopoulou (National Technical University of Athens) and Francesca Belloni (CEA Saclay) for answering to our questions on their experimental data.
EXFOR compilation of the $^{232}$Th/$^{235}$U ratios measured at the CSNS Back-n and CERN n\_TOF facilities were done with Jimin Wang (China Institute of Atomic Energy) and Emmeric Dupont (CEA Saclay), respectively.
We also would like to thank the members of the International Network of Nuclear Reaction Data Centres (NRDC) for maintenance and development of the EXFOR library.
\bibliography{vid-jnst.bib}
\end{document}